\documentclass{ws-ijmpa}

\usepackage{amsthm}
\usepackage{amsfonts}
\usepackage{xspace}

\usepackage{graphicx}
\usepackage{multirow}
\usepackage{subfigure}
\usepackage{lineno}
\usepackage{hyperref}
\usepackage[normalem]{ulem}
\hypersetup{
	colorlinks=true,
	linkcolor=blue,
	citecolor=blue,
	urlcolor=blue
}
\newcommand{\customUL}{\bgroup\markoverwith{\textcolor{blue}{\rule[-0.3ex]{2pt}{0.6pt}}}\ULon}

\usepackage[super,compress]{cite}
\usepackage{graphicx}
\usepackage{booktabs}

\usepackage{lineno}

\newcommand{\AtlasCoordFootnote}{%
	ATLAS uses a right-handed coordinate system,
	with its origin at the nominal interaction point (IP) in the centre of the detector and the \(z\)-axis along the beam pipe. The \(x\)-axis points from the IP to the centre of the LHC ring, and the \(y\)-axis points upwards. Cylindrical coordinates \((r,\phi)\) are used in the transverse plane,  \(\phi\) being the azimuthal angle around the \(z\)-axis.
	The pseudorapidity is defined in terms of the polar angle \(\theta\) as \(\eta = -\ln \tan(\theta/2)\).
	Angular distance is measured in units of \(\Delta R \equiv \sqrt{(\Delta\eta)^{2} + (\Delta\phi)^{2}}\).}

\newcommand{\gev}{~\ensuremath{\mathrm{GeV}}\xspace}

\newcommand{\pt}{\ensuremath{p_T}\xspace}
\newcommand{\et}{\ensuremath{E_T}\xspace}

\def\antibar#1{\ensuremath{#1\bar{#1}}}

\def\ttbar{\antibar{t}}

\def\ggino{\ensuremath{\mathchoice%
      {\displaystyle\raise.4ex\hbox{$\displaystyle\tilde\chi$}}%
         {\textstyle\raise.4ex\hbox{$\textstyle\tilde\chi$}}%
       {\scriptstyle\raise.3ex\hbox{$\scriptstyle\tilde\chi$}}%
 {\scriptscriptstyle\raise.3ex\hbox{$\scriptscriptstyle\tilde\chi$}}}}

\def\chinop{\ensuremath{\mathchoice%
      {\displaystyle\raise.4ex\hbox{$\displaystyle\tilde\chi^+$}}%
         {\textstyle\raise.4ex\hbox{$\textstyle\tilde\chi^+$}}%
       {\scriptstyle\raise.3ex\hbox{$\scriptstyle\tilde\chi^+$}}%
 {\scriptscriptstyle\raise.3ex\hbox{$\scriptscriptstyle\tilde\chi^+$}}}}
\def\chinom{\ensuremath{\mathchoice%
      {\displaystyle\raise.4ex\hbox{$\displaystyle\tilde\chi^-$}}%
         {\textstyle\raise.4ex\hbox{$\textstyle\tilde\chi^-$}}%
       {\scriptstyle\raise.3ex\hbox{$\scriptstyle\tilde\chi^-$}}%
 {\scriptscriptstyle\raise.3ex\hbox{$\scriptscriptstyle\tilde\chi^-$}}}}
\def\chinopm{\ensuremath{\mathchoice%
      {\displaystyle\raise.4ex\hbox{$\displaystyle\tilde\chi^\pm$}}%
         {\textstyle\raise.4ex\hbox{$\textstyle\tilde\chi^\pm$}}%
       {\scriptstyle\raise.3ex\hbox{$\scriptstyle\tilde\chi^\pm$}}%
 {\scriptscriptstyle\raise.3ex\hbox{$\scriptscriptstyle\tilde\chi^\pm$}}}}
\def\chinomp{\ensuremath{\mathchoice%
      {\displaystyle\raise.4ex\hbox{$\displaystyle\tilde\chi^\mp$}}%
         {\textstyle\raise.4ex\hbox{$\textstyle\tilde\chi^\mp$}}%
       {\scriptstyle\raise.3ex\hbox{$\scriptstyle\tilde\chi^\mp$}}%
 {\scriptscriptstyle\raise.3ex\hbox{$\scriptscriptstyle\tilde\chi^\mp$}}}}

\def\chinoonep{\ensuremath{\mathchoice%
      {\displaystyle\raise.4ex\hbox{$\displaystyle\tilde\chi^+_1$}}%
         {\textstyle\raise.4ex\hbox{$\textstyle\tilde\chi^+_1$}}%
       {\scriptstyle\raise.3ex\hbox{$\scriptstyle\tilde\chi^+_1$}}%
 {\scriptscriptstyle\raise.3ex\hbox{$\scriptscriptstyle\tilde\chi^+_1$}}}}
\def\chinoonem{\ensuremath{\mathchoice%
      {\displaystyle\raise.4ex\hbox{$\displaystyle\tilde\chi^-_1$}}%
         {\textstyle\raise.4ex\hbox{$\textstyle\tilde\chi^-_1$}}%
       {\scriptstyle\raise.3ex\hbox{$\scriptstyle\tilde\chi^-_1$}}%
 {\scriptscriptstyle\raise.3ex\hbox{$\scriptscriptstyle\tilde\chi^-_1$}}}}
\def\chinoonepm{\ensuremath{\mathchoice%
      {\displaystyle\raise.4ex\hbox{$\displaystyle\tilde\chi^\pm_1$}}%
         {\textstyle\raise.4ex\hbox{$\textstyle\tilde\chi^\pm_1$}}%
       {\scriptstyle\raise.3ex\hbox{$\scriptstyle\tilde\chi^\pm_1$}}%
 {\scriptscriptstyle\raise.3ex\hbox{$\scriptscriptstyle\tilde\chi^\pm_1$}}}}

\def\chinotwop{\ensuremath{\mathchoice%
      {\displaystyle\raise.4ex\hbox{$\displaystyle\tilde\chi^+_2$}}%
         {\textstyle\raise.4ex\hbox{$\textstyle\tilde\chi^+_2$}}%
       {\scriptstyle\raise.3ex\hbox{$\scriptstyle\tilde\chi^+_2$}}%
 {\scriptscriptstyle\raise.3ex\hbox{$\scriptscriptstyle\tilde\chi^+_2$}}}}
\def\chinotwom{\ensuremath{\mathchoice%
      {\displaystyle\raise.4ex\hbox{$\displaystyle\tilde\chi^-_2$}}%
         {\textstyle\raise.4ex\hbox{$\textstyle\tilde\chi^-_2$}}%
       {\scriptstyle\raise.3ex\hbox{$\scriptstyle\tilde\chi^-_2$}}%
 {\scriptscriptstyle\raise.3ex\hbox{$\scriptscriptstyle\tilde\chi^-_2$}}}}
\def\chinotwopm{\ensuremath{\mathchoice%
      {\displaystyle\raise.4ex\hbox{$\displaystyle\tilde\chi^\pm_2$}}%
         {\textstyle\raise.4ex\hbox{$\textstyle\tilde\chi^\pm_2$}}%
       {\scriptstyle\raise.3ex\hbox{$\scriptstyle\tilde\chi^\pm_2$}}%
 {\scriptscriptstyle\raise.3ex\hbox{$\scriptscriptstyle\tilde\chi^\pm_2$}}}}

\def\nino{\ensuremath{\mathchoice%
      {\displaystyle\raise.4ex\hbox{$\displaystyle\tilde\chi^0$}}%
         {\textstyle\raise.4ex\hbox{$\textstyle\tilde\chi^0$}}%
       {\scriptstyle\raise.3ex\hbox{$\scriptstyle\tilde\chi^0$}}%
 {\scriptscriptstyle\raise.3ex\hbox{$\scriptscriptstyle\tilde\chi^0$}}}}

\def\ninoone{\ensuremath{\mathchoice%
      {\displaystyle\raise.4ex\hbox{$\displaystyle\tilde\chi^0_1$}}%
         {\textstyle\raise.4ex\hbox{$\textstyle\tilde\chi^0_1$}}%
       {\scriptstyle\raise.3ex\hbox{$\scriptstyle\tilde\chi^0_1$}}%
 {\scriptscriptstyle\raise.3ex\hbox{$\scriptscriptstyle\tilde\chi^0_1$}}}}
\def\ninotwo{\ensuremath{\mathchoice%
      {\displaystyle\raise.4ex\hbox{$\displaystyle\tilde\chi^0_2$}}%
         {\textstyle\raise.4ex\hbox{$\textstyle\tilde\chi^0_2$}}%
       {\scriptstyle\raise.3ex\hbox{$\scriptstyle\tilde\chi^0_2$}}%
 {\scriptscriptstyle\raise.3ex\hbox{$\scriptscriptstyle\tilde\chi^0_2$}}}}
\def\ninothree{\ensuremath{\mathchoice%
      {\displaystyle\raise.4ex\hbox{$\displaystyle\tilde\chi^0_3$}}%
         {\textstyle\raise.4ex\hbox{$\textstyle\tilde\chi^0_3$}}%
       {\scriptstyle\raise.3ex\hbox{$\scriptstyle\tilde\chi^0_3$}}%
 {\scriptscriptstyle\raise.3ex\hbox{$\scriptscriptstyle\tilde\chi^0_3$}}}}
\def\ninofour{\ensuremath{\mathchoice%
      {\displaystyle\raise.4ex\hbox{$\displaystyle\tilde\chi^0_4$}}%
         {\textstyle\raise.4ex\hbox{$\textstyle\tilde\chi^0_4$}}%
       {\scriptstyle\raise.3ex\hbox{$\scriptstyle\tilde\chi^0_4$}}%
 {\scriptscriptstyle\raise.3ex\hbox{$\scriptscriptstyle\tilde\chi^0_4$}}}}

\newcommand*{\etcx}{\ensuremath{E_{\mathrm{T}}^{\mathrm{coneXX}}}\xspace}
\newcommand*{\etcs}{\ensuremath{E_{\mathrm{T}}^{\mathrm{cone20}}}\xspace}

\newcommand*{\ptcs}{\ensuremath{p_{\mathrm{T}}^{\mathrm{cone20}}}\xspace}
\newcommand*{\ptvcs}{\ensuremath{p_{\mathrm{T}}^{\mathrm{varcone20}}}\xspace}
\newcommand*{\ptcx}{\ensuremath{p_{\mathrm{T}}^{\mathrm{coneXX}}}\xspace}

\newcommand{\muhat}{\ensuremath{\left\langle \mu \right\rangle}}

\newcommand{\cellsig}{\ensuremath{\varsigma_\mathrm{cell}^\mathrm{EM}}}

\begin{document}
	
	
\markboth{Otilia Ducu}{Electron efficiency in LHC Run-2 with the ATLAS experiment}

%
%

\title{Electron efficiency in LHC Run-2 with the ATLAS experiment 
}

\author{Otilia Ducu\\ (on behalf of the ATLAS Collaboration)
}

\address{Horia Hulubei National Institute of Physics and Nuclear Engineering (IFIN-HH)\\
Magurele, Ilfov, Romania, 077125\\
otilia.ducu@gmail.com}

\maketitle


\begin{abstract}
The document presents a general overview of the electron reconstruction, identification and isolation performance in the ATLAS experiment. The results are obtained using 13~TeV proton-proton collision data collected during the LHC Run-2. The electron reconstruction efficiency is higher than 97\%, and the ratio of data to Monte Carlo simulation efficiency is close to unity, with associated uncertainties generally smaller than 0.1\%. The electron identification is shown for three working points, and depending on the electron $E_T$, it can be as low as 60\%, increasing to more than 80\% above 50~GeV. The correction factors are close to one, generally within 5\%.
Five isolation working points are recommended in the ATLAS experiment, to successfully reject fake/non-prompt electrons. Their dependency on the electron identification working points is shown and discussed, as well as their pile-up dependency, and their performance versus electron $E_T$ and $\eta$.

\textit{Document based on a presentation at the XI International Conference on New Frontiers in Physics (ICNFP 2022).}

\keywords{prompt electrons, reconstruction, identification, isolation, fake/non-prompt electrons}
\end{abstract}

\ccode{PACS numbers: 06.30.-k, 14.60.-z}


%
%

\section{Introduction}
The ATLAS experiment has made significant contributions to the Higgs boson physics, both in the discovery~\cite{Higgs_discovery_ATLAS,Higgs_discovery_CMS} and the precision measurements of its properties. The first observations~\cite{Higgs_discovery_ATLAS} were in the $H\to ZZ^* \to 4\ell$ ($\ell$ = $e$, $\mu$), $H \to \gamma\gamma$ and $H\to W W^* \to e\nu \mu\nu$ channels, which required the excellent performance and understanding of the detector and its components: the tracker, the electromagnetic and hadronic calorimeters. The photon and electron reconstruction algorithm also played a crucial role in this achievement. However, the discovery of the Higgs boson was not the end of the story. Since then, more and more effort was put into the precision measurements of the Higgs boson related parameters, such as its mass, couplings, and width. These measurements, or more generally all the precision Standard Model measurements, rely on understanding the systematic uncertainties of the object reconstruction in the detector with high precision.

This document presents a general overview of the electron reconstruction, identification and isolation efficiencies measurements performed with the ATLAS experiment. The results were obtained with the Run-2 data from the Large Hadron Collider (LHC), and used for most of the recent ATLAS precision measurements of the Standard Model parameters~\cite{SM_PM_1,SM_PM_2,SM_PM_3}.

\section{LHC and ATLAS detector}

\begin{figure}[!tb]
	\begin{center}
			\includegraphics[width=0.78\columnwidth]{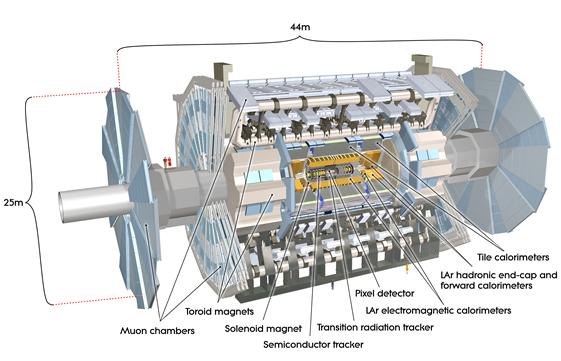}
	\end{center}
	\caption{
		Illustration of the ATLAS detector. Reused with permission from Ref.~\citen{ATLAS_det_fig}.
	}
	\label{fig:ATLAS_det_fig}
\end{figure}

The LHC is the largest and most powerful hadron collider in the world, a research program approved in December 1994~\cite{LHC_1,LHC_2}. The proton-proton (pp) Run-2 data taking started during 2015 and finished in 2018, and in this period operated at an energy in the center of mass of 13 TeV. ATLAS and CMS are the general-purpose particle physics  experiments at the LHC, and both were designed and constructed to achieve similar physics goals, but using different technologies. The ATLAS detector covers nearly the entire solid angle around the collision point\footnote{\AtlasCoordFootnote}, and has four main sub-systems~\cite{ATLAS_1,ATLAS_2}, as illustrated in Figure~\ref{fig:ATLAS_det_fig}. 

The ATLAS experiment relies on the inner detector for the tracking of charged particles coming from the $pp$ collisions. It is closest to the interaction point, surrounded by a 2~T magnetic field, and covers the pseudorapidity range of $|\eta|<2.5$. The inner detector consists of three sub-detectors: the pixel detector, the silicon microstrip tracker (SCT) and the transition radiation tracker (TRT). They ensure an accurate reconstruction and identification of tracks from primary and secondary vertices, the latter permitting an efficient reconstruction of photon conversions in the inner detector. The TRT sub-detector helps to improve the electron identification, through the detection of transition radiation photons. 

The ATLAS calorimeter system covers the $|\eta|<4.9$ range, and is formed by the LAr electromagnetic (EM), and the hadronic calorimeters. The EM calorimeter stops and determines the energy deposited by electrons and photons, and provides all the information necessary for a precise electron and photon reconstruction, and identification. It is split in two half-barrels that cover the $|\eta|< 1.475$ range, and two end-cap coaxial wheels that cover the $1.375 < |\eta|< 3.2$ range. 
The transition region between the barrel and the end-cap, or the so-called \textit{crack region}, is in the $1.37 < |\eta| < 1.52$ range and has a relatively large amount of inactive material.
The EM calorimeter has liquid argon as active medium, with lead and stainless steel absorbers (and copper electrodes) that have an accordion shape. This geometry and material composition allows a high granularity, and a high $\eta - \varphi$ resolution. This also ensures large EM shower shapes for the charged particles at the passage through the calorimeter, key signatures that help to distinguish the real electron from e.g jets faking electrons.
In the $|\eta|<2.47$ region (excluding the crack region), the EM calorimeter is divided into three longitudinal compartments called the first, second, and third layers. 
To correct for energy loss in material upstream of the calorimeters, the LAr EM calorimeter is preceded by an additional thin LAr presampler (that does not have absorbers) covering the $|\eta| < 1.8$ region.

The hadronic calorimeter surrounds the EM calorimeter, and is composed by tile, LAr hadronic end-caps and LAr forward calorimeters. It has a less fine granularity than the EM calorimeter, and is responsible for the jet reconstruction and transverse momentum computation.

\begin{figure}[!tb]
\begin{center}
		\includegraphics[width=0.88\columnwidth]{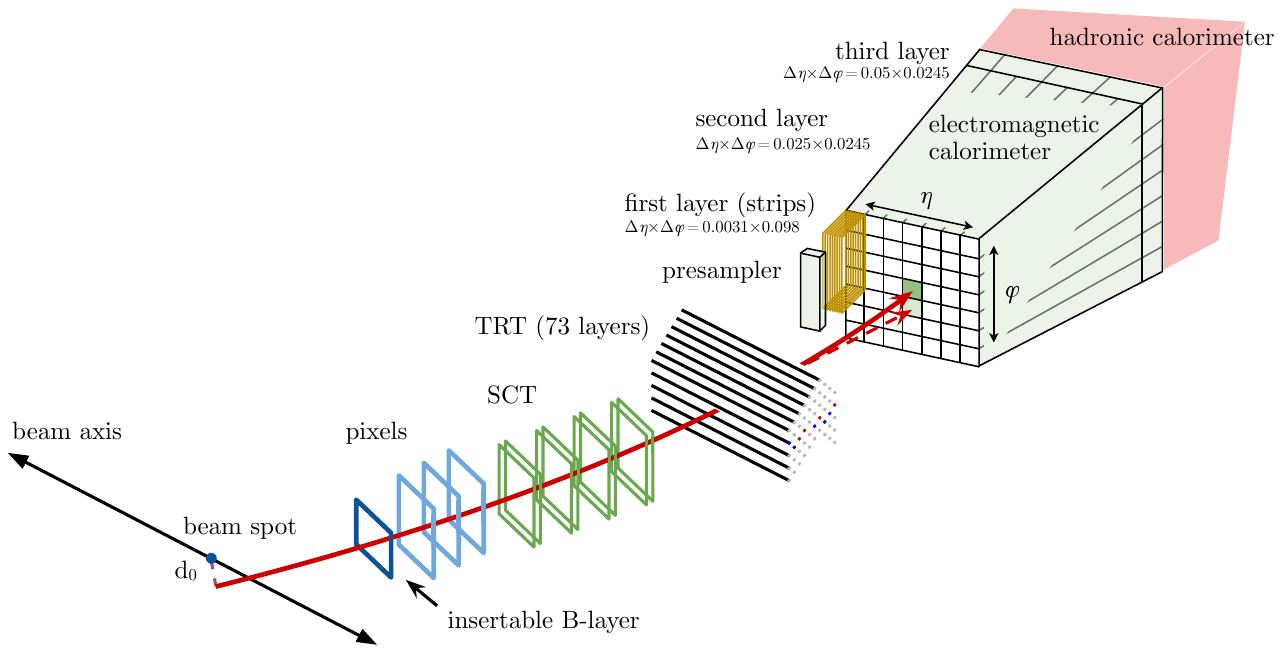}
\end{center}
\caption{
	Illustration of the electron path through the ATLAS sub-detectors. The red, hashed line illustrates the path of a photon produced by the interaction of the electron with the material in the tracking system.
	Reused with permission from Ref.~\citen{Egamma_201516}.
}
\label{fig:ElePath}
\end{figure}
The path of an electron through the various sub-systems of the ATLAS inner detector and calorimeters discussed above, is illustrated in Figure~\ref{fig:ElePath}. The figure shows also the $\Delta \eta \times \Delta \varphi$ granularity of the three layers composing the barrel EM calorimeter.

Unlike ATLAS, the CMS detector~\cite{CMS_det} uses lead tungstate scintillating crystals for the EM calorimeter. The crystals are very dense, almost transparent and emit light when electrons and photons pass through them. This light is captured by photodetectors attached to the crystals, and converted into an electrical signal that is amplified and further analyzed. The EM calorimeter has a barrel and two end-cap regions, like ATLAS. The advantage of using these crystals is that they produce light in quick, short, and well-defined photon bursts, which enable a fast, accurate, and compact detector.
To separate the potentially interesting single high-energy photons from the less interesting close pairs of low-energy photons (for example, from $\pi^0$ decays), there is a preshower detector in front of the EM calorimeter end-caps. As it will be shown later, the performance of electron and photon reconstruction and identification is equally impressive in CMS, as in ATLAS.

The ATLAS hadronic calorimeter surrounds the EM calorimeter, and has a poorer resolution than the latter. It measures the energy deposited by the hadronic particles that do not deposit all their energy in the EM calorimeter. It has three sub-systems: the steel/scintillator Tile, the copper/LAr hadronic end-caps, and the copper/LAr forward calorimeters. The tile calorimeter covers the $|\eta|<1.7$ region, and the hadronic end-caps cover the $1.5<|\eta|<3.2$ region, respectively. The forward calorimeters extend the coverage up to $|\eta| = 4.9$.

The muon spectrometer is the last sub-detector, the envelope of the ATLAS detector, as illustrated in Figure~\ref{fig:ATLAS_det_fig}. It has a dedicated trigger, and is composed of a magnetic system and several high precision tracking chambers organized in stations: one in the barrel, and two in the end-caps. It covers the $|\eta|<2.7$ range, with the muon trigger system only covering the  $|\eta| < 2.4$ range. The muon spectrometer is responsible for measuring the momentum and identifying muon tracks with a high precision.

The interesting events are first selected by the custom hardware-based first-level trigger system. Then, the software-based algorithms in the high-level trigger system make further selections~\cite{ATLAS:2016wtr}. To obtain the results presented in this document, the lowest unprescaled triggers were used. 

An extensive software suite~\cite{ATL-SOFT-PUB-2021-001} was used in data simulation, in the reconstruction and analysis of real and simulated data, in detector operations, and in the trigger and data acquisition systems of the experiment.

\section{Electron reconstruction}

\begin{figure}[!tb]
\begin{center}
	\includegraphics[width=0.75\columnwidth]{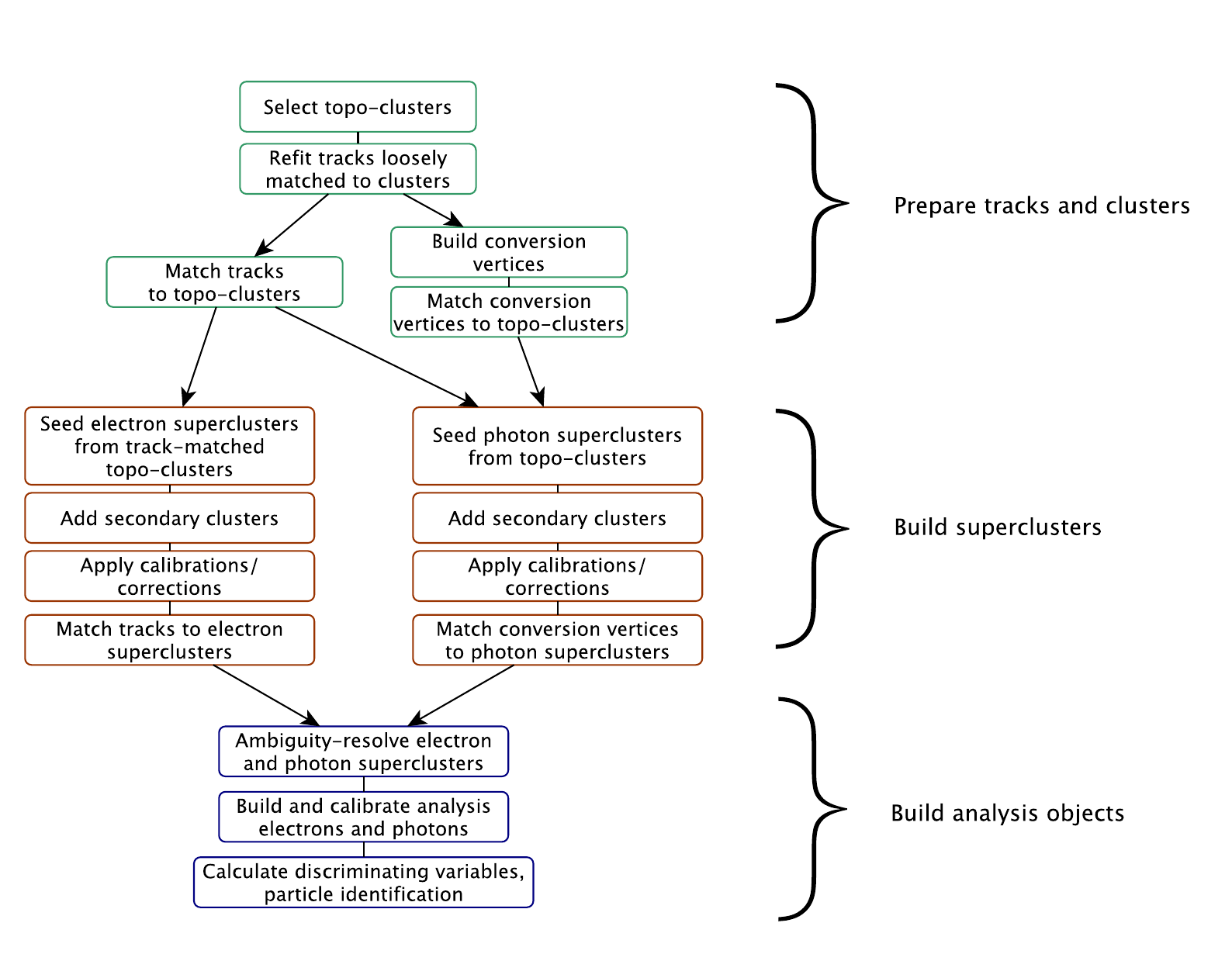}
\end{center}
\caption{	
	Summary of various steps performed for the electron and photon reconstruction in ATLAS, in the barrel region of the detector. 
	Reused with permission from Ref.~\citen{Egamma_201517}.
}
\label{fig:ElePh_Reco}
\end{figure}

The reconstruction of electrons and photons in the ATLAS experiment is based on dynamic, variable-size clusters of calorimeter cells, also called superclusters (SCs). These SCs can capture better the energy from bremsstrahlung photons, or from electrons that originate from photon conversions. This procedure is explained in detail in Refs.~\refcite{Egamma_201516} and~\refcite{Egamma_201517}, and only a brief overview is given here. Figure~\ref{fig:ElePh_Reco} shows a diagram of the various steps performed during the reconstruction process.

The algorithm prepares the topo-clusters and the tracks first. The topo-clusters reconstruction starts with calorimeter cells that have a significance $\cellsig (= \left| \frac{E_\text{cell}^\text{EM}}{\sigma_\text{noise,cell}^\text{EM}} \right|)$ of at least 4. These cells form the initial clusters, or the proto-clusters. Then, the neighboring cells that have $\cellsig$ of at least 2 are added to the proto-clusters. These cells also become the seeds for the next iteration that collects their neighbors in the proto-cluster. This process continues until all the nearby cells are included in the proto-cluster. Finally, the neighboring cells that have $\cellsig \geq 0$ are added to the cluster. Note that the electron (and photon) reconstruction starts from the topo-clusters, but only considers the energy from cells in the EM calorimeter.

A standard track-pattern reconstruction algorithm is used to perform the electron track reconstruction. It uses the hit information in the Pixel and SCT detectors to form clusters. The track reconstruction has three steps: pattern recognition, ambiguity resolution, and TRT extension. The pattern recognition uses the pion hypothesis to model the energy loss from particle interactions with the detector material. The track candidates are fitted with the ATLAS Global $\chi^2$ fitter, and the tracks that have hits in the SCT loosely matched to the EM clusters are re-fitted with a Gaussian sum filter (GSF) algorithm. This procedure helps to better account for the electron energy loss in the detector material.

Using these two inputs, the algorithm matches the tracks and the topo-clusters. In parallel, it also builds the conversion vertices and connects them to the topo-clusters. It proceeds with the electron and photon supercluster-building steps, that run separately. Once the superclusters are build, an initial energy calibration and position correction is applied. Next, the algorithm connects the tracks to the electron superclusters, and the conversion vertices to the photon superclusters. The final step is building and calibrating the analysis-level electron (and photon) objects -- it also resolves the ambiguity between electrons and photons.

\begin{figure}[!tb]
	\begin{center}
		\includegraphics[width=0.45\columnwidth]{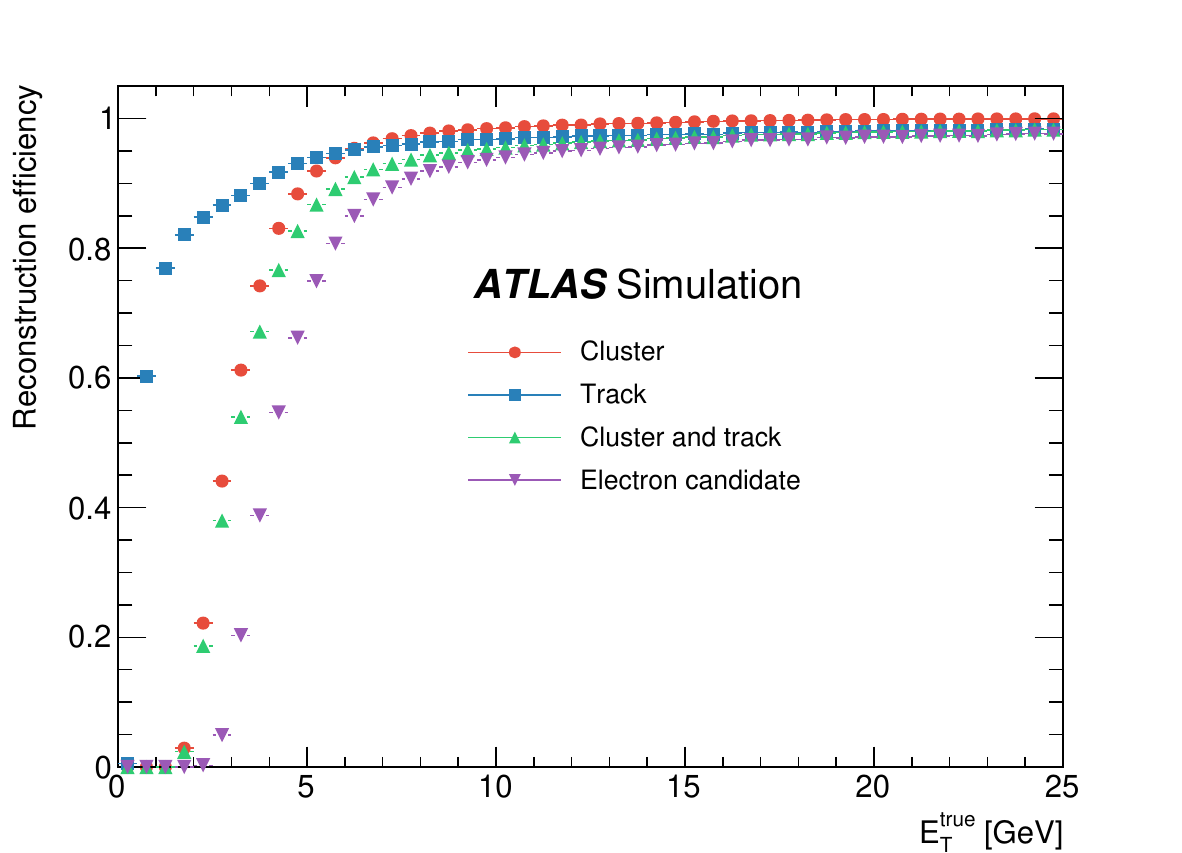}
		\includegraphics[width=0.42\columnwidth]{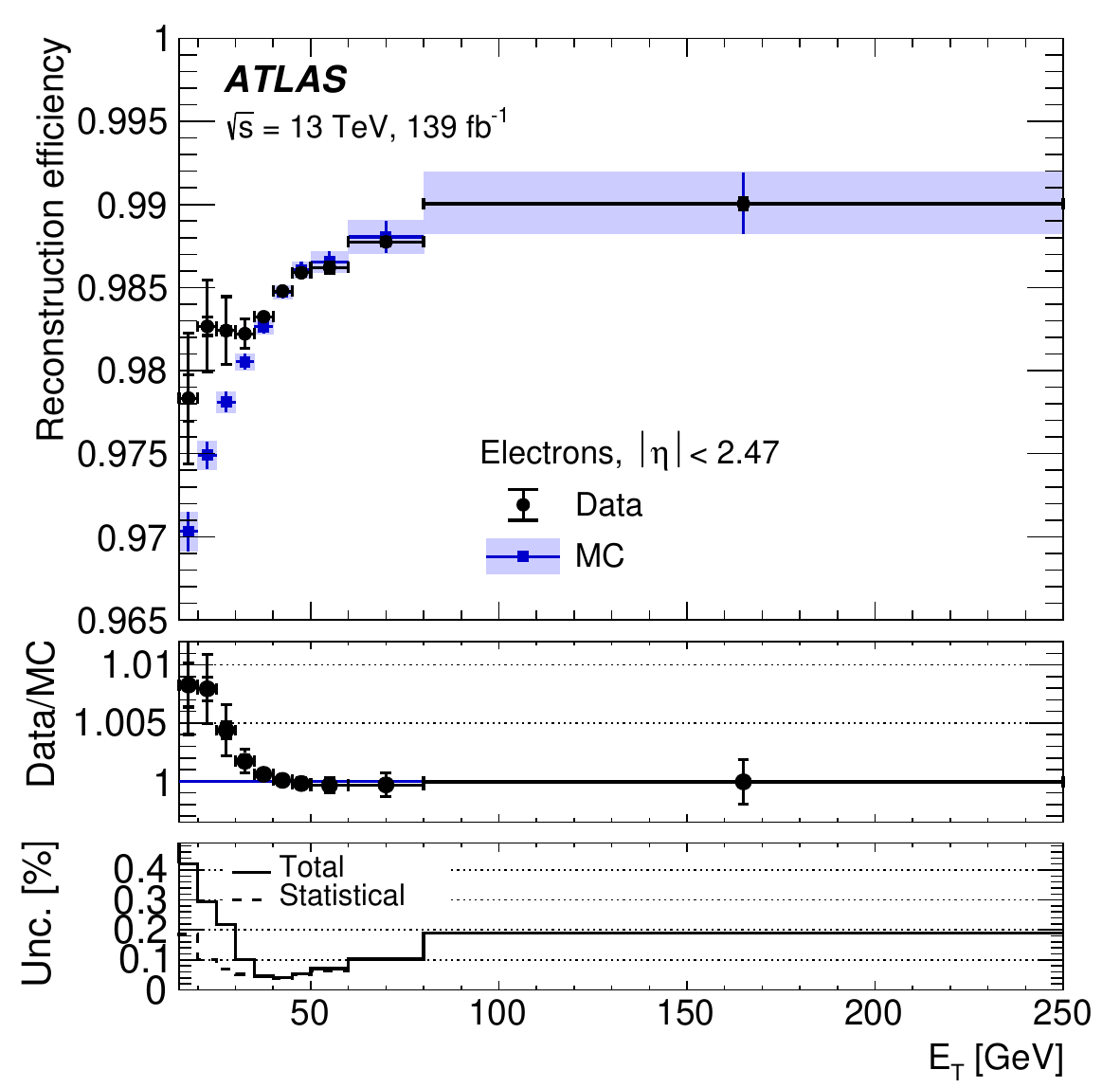}
	\end{center}
	\caption{	
		Electron reconstruction efficiency.
		Reused with permission from Refs.~\citen{Egamma_201517,Egamma_2023}.
	}
	\label{fig:Ele_Reco_Eff}
\end{figure}

To illustrate the performance of the various steps in the reconstruction algorithm, some rough measurements of reconstruction efficiencies for the cluster, the track, the matched-cluster track, and the electron candidate are shown in Figure~\ref{fig:Ele_Reco_Eff}, left, as a function of the generated (true) electron transverse momentum, $E_T$, in MC simulation. The efficiencies are always above 95\%, when $E_T > 10$~GeV. Below 5~GeV they drop significantly, as the electron reconstruction is more challenging, given the detector limitations, and the high amount of background and pile-up.

As mentioned in the introduction, a precise measurement of the electron reconstruction efficiency in data and in MC simulation is crucial. In data, the large amount of background, as well as the available triggers, makes impossible a measurement in the $E_T < 15$~GeV range. Above 15~GeV, the measurement is performed using the Tag\&Probe method presented in Ref.~\citen{Egamma_2012}, in an $Z \to ee$ enriched region.
The backgrounds are from electrons with an associated track, and from electrons with no associated track (that are reconstructed as photons).
They are more present in the $E_T < 30$~GeV range, and overall estimated using data-based techniques.
Figure~\ref{fig:Ele_Reco_Eff} right, shows the efficiencies obtained using the LHC Run-2 data and in $Z\to ee$ MC simulation. The middle pad shows the data over MC simulation ratio, while the bottom pad shows the total and statistical uncertainties associated with this ratio.
The electron reconstruction efficiency is above 97\%, but it drops at higher $|\eta|$ and in the calorimeter transition region~\cite{Egamma_2023}. Similar values are also measured in CMS~\cite{CMS_Egamma}.

The results actually used by the ATLAS analyses are the data over MC simulation efficiency ratios, or the so-called correction factors. These are applied as event weights to correct the MC simulation, in order to reproduce as best as possible the electron reconstruction efficiency in data. As shown in Figure~\ref{fig:Ele_Reco_Eff} right, these are close to unity. Below 30~GeV in $E_T$, these correction factors are as high as 1.01, as in this region the background is much higher and more challenging to estimate. The associated systematic uncertainties are generally less than 0.5\% (0.1\%) when $E_T < 30$~GeV ($E_T > 30$~GeV).
Below 15~GeV the correction factors are assumed to be equal to 1~$\pm$2\% (5\%) in the barrel (end-cap) region. The correction factors could be dependent on pileup, and this effect was studied and found to be negligible. In CMS, the reconstruction efficiency is claimed to be compatible between data and simulation within 2\%.

\section{Electron identification}
The reconstructed electrons can be isolated/prompt, or fake/non-prompt. To remove the unwanted fake/non-prompt electrons, 3 selection working points are defined using a likelihood based identification: Loose, Medium and Tight, ordered with increasing background rejection power~\cite{Egamma_201516}. The likelihood (LLH) discriminant is constructed from quantities measured only in the inner detector or only in the calorimeter, or using the combined inner detector and calorimeter information. These variables (see Table 1 in Ref.~\citen{Egamma_201516}) can discriminate prompt electrons from energy deposits from hadronic jets and converted photons, or from non-prompt electrons produced in heavy flavor hadrons decays. The track variables are required to satisfy a set of quality requirements, and all EM shower shape variables are calculated by summing energy deposits in calorimeter cells relative to the cluster’s hottest cell~\cite{Egamma_2023}.

\begin{figure}[!tb]
\begin{center}
	\includegraphics[width=0.42\columnwidth]{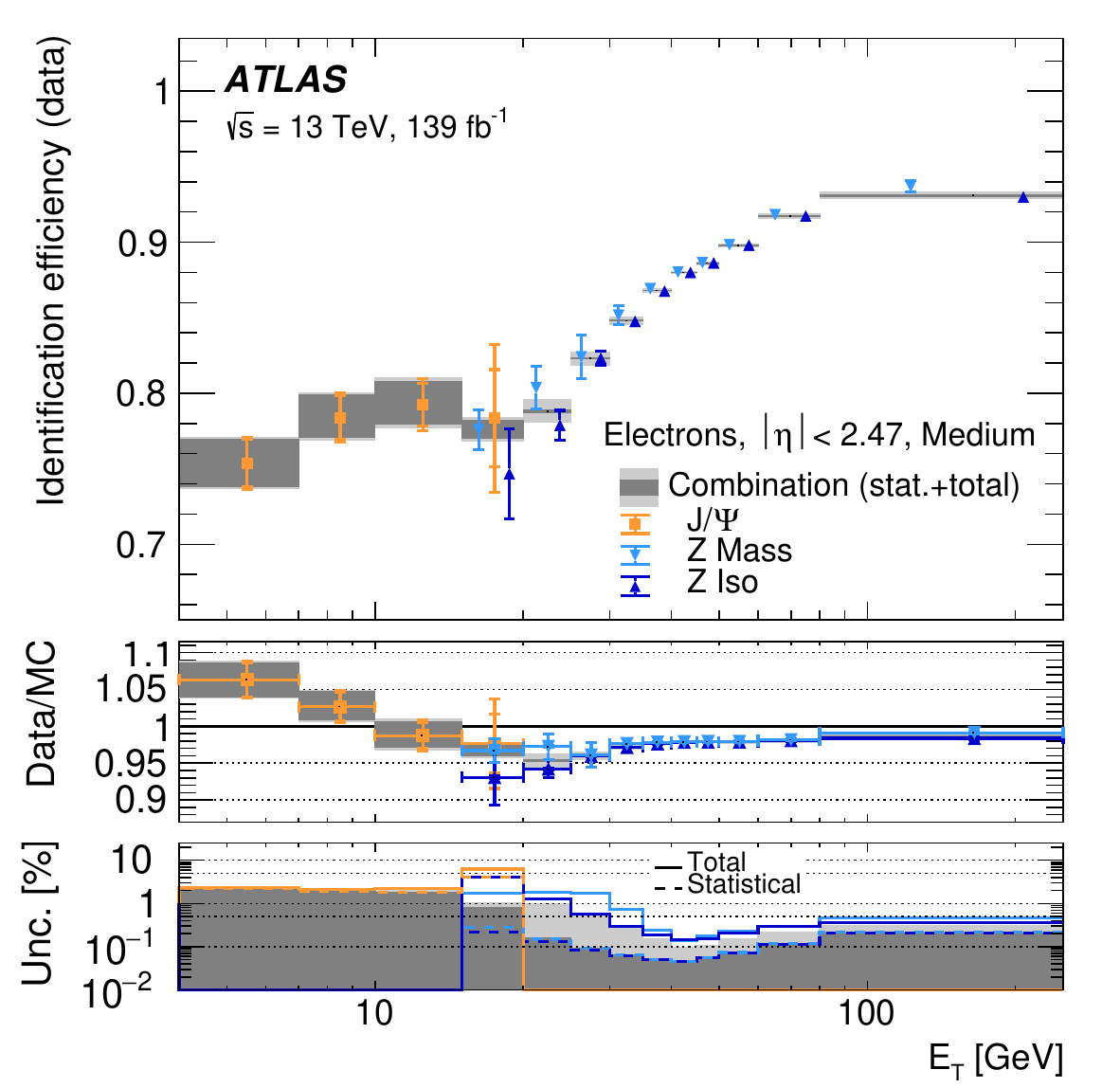}	
	\includegraphics[width=0.42\columnwidth]{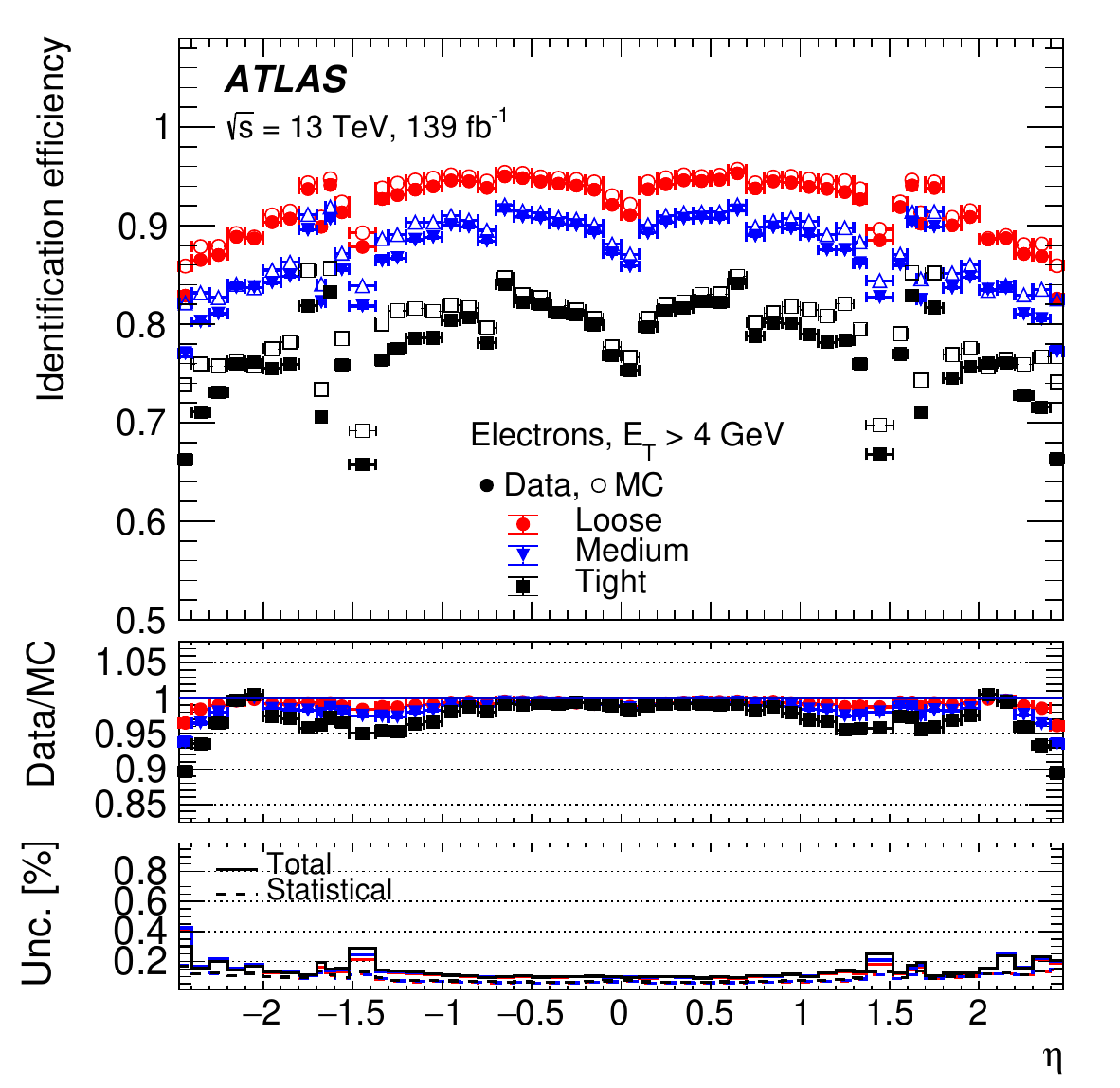}
	\includegraphics[width=0.42\columnwidth]{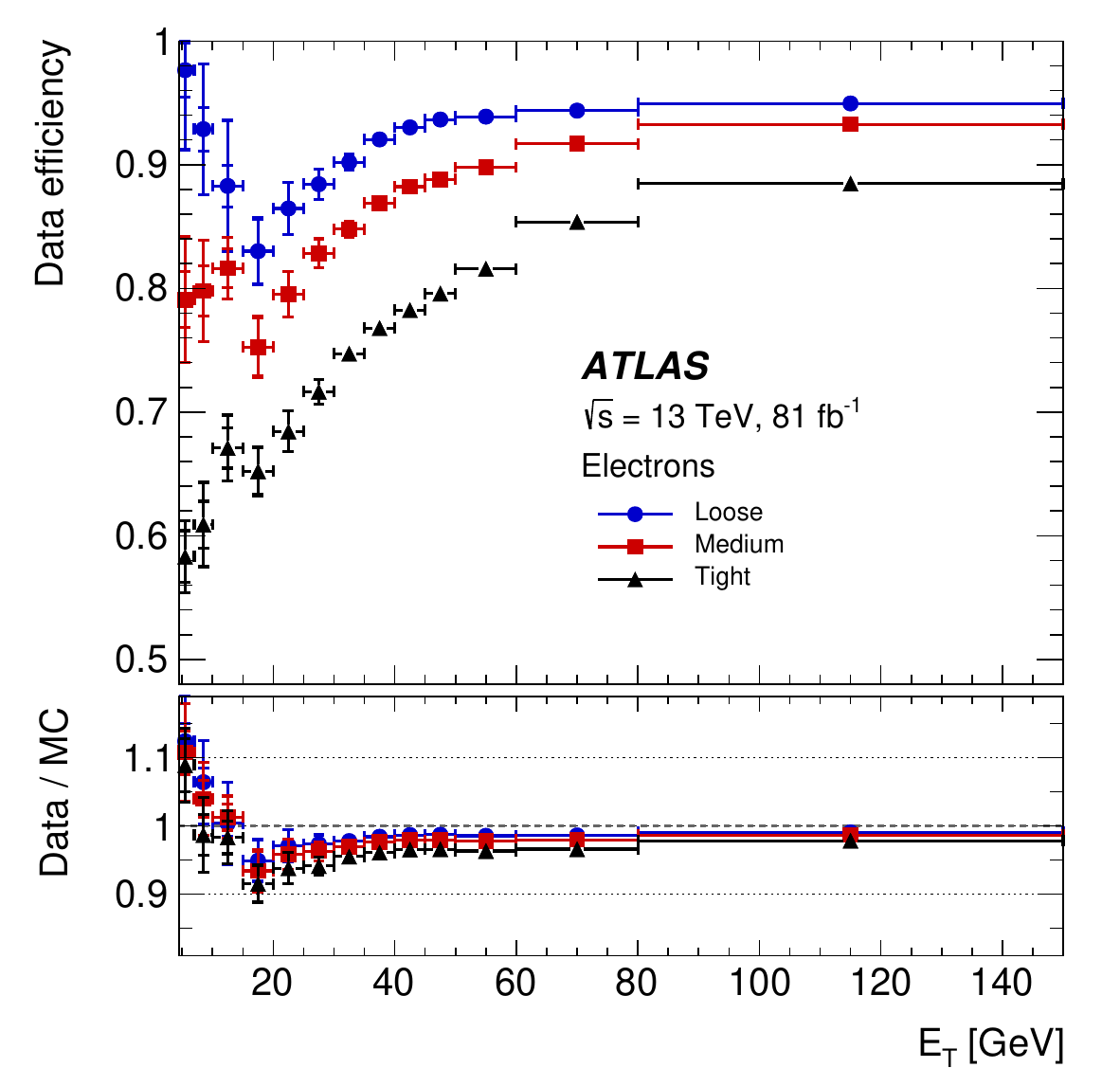}
	\includegraphics[width=0.42\columnwidth]{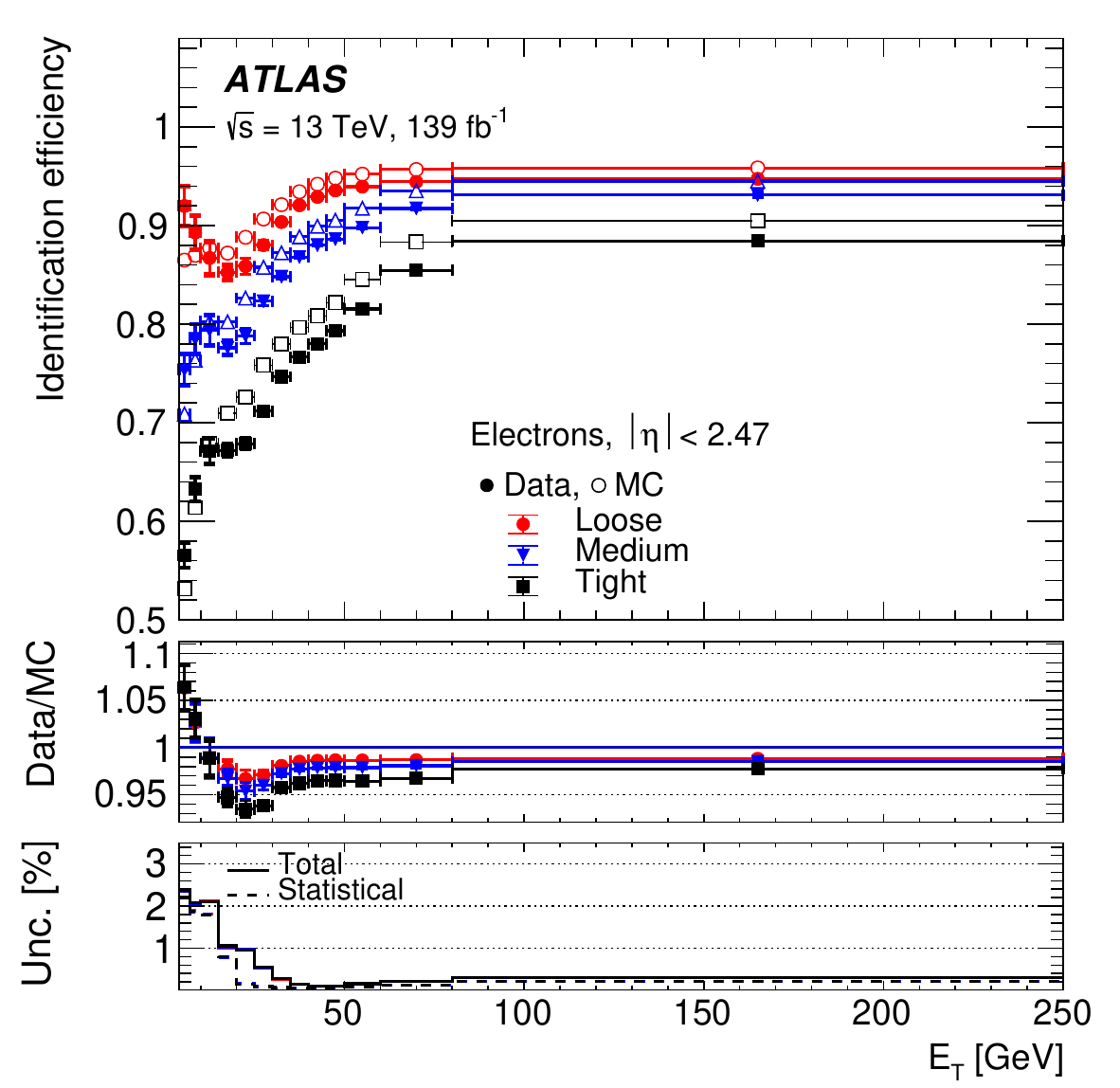}
\end{center}
\caption{	
	Electron identification efficiency. 
	Reused with permission from Refs.~\citen{Egamma_201517,Egamma_2023}.
}
\label{fig:Ele_ID_Eff}
\end{figure}

To measure the efficiency of the electron identification working points, the Tag\&Probe method is used, in 234 bins of electron $E_T$ and $\eta$. This fine binning is necessary to account for the detector geometry, interaction effects with the detector material, as well as process kinematics. The measurements in the $4.5 < E_T < 20$~GeV region are performed using $J/\Psi \to ee$ events, and in the $15 < E_T < 200$~GeV region using $Z \to ee$ events, respectively. The $J/\Psi\to ee$ measurement uses the invariant-mass distribution of the two electron candidates and exploits the pseudo-proper time variable, while the $Z \to ee$ measurement uses either the invariant-mass distribution ($Z$-mass) or the amount of transverse energy in an isolation cone around the probe electron ($Z$-isolation)~\cite{Egamma_2023}. The most challenging task, for all methods, is the precise estimation of the background. In the overlapping region, the results from the two independent measurements are combined. The gain from the combination is clearly seen in Figure~\ref{fig:Ele_ID_Eff}, top-left, as a massive reduction of the uncertainty on the final efficiency, and thus on the correction factors.

Figure~\ref{fig:Ele_ID_Eff}, top-right and bottom, shows the electron identification efficiency for the 3 working points, in $Z\to ee$ data and MC simulation. The middle panel shows the data over MC simulation ratio, and the bottom panel shows the relative uncertainties, respectively. The identification efficiency for the Loose working point is around 80\% for electrons with $E_T < 40$~GeV, and increases to 93\% above 80~GeV. The efficiency for Tight is as low at 60\% below 15~GeV, increasing to 75\% (88\%) for electrons with $E_T$ around 40~GeV ($>100$~GeV). The performance of the Medium working point is between the Loose and Tight ones, as designed. The electron identification efficiencies are symmetric in the positive and negative $\eta$ regions of the detector. The correction factors for all three electron identification working points are close to one, within 5\%. A similar performance is obtained also in the CMS experiment~\cite{CMS_Egamma}.

Figure~\ref{fig:Ele_ID_Eff}, bottom, shows two sets of measurements versus $E_T$: using 81~fb$^{-1}$ of data (left), and 139~fb$^{-1}$ of data (right). Most interesting is to compare the results in low $E_T$ region, let us say below 20-25~GeV. The 139~fb$^{-1}$ results are using an improved methodology for the background estimation, for the $Z$-mass method~\cite{Egamma_2023}. This work has greatly improved the accuracy of the efficiencies and the correction factors. The error in the efficiency has gone down by about 30\% for the Loose working point, and by more than 50\% for the Medium and Tight working points. The improvement is similar for all values of $\eta$. Note that the drop in the efficiency around 15~GeV is known, and caused by a mismodeling of variables used in the likelihood discriminant at low $E_T$ (more details are in Ref.~\citen{Egamma_201517}).

\section{Electron isolation}
In addition to the electron identification working points, isolation requirements are applied to further remove the fake/non-prompt electrons. A total of 5 isolation working points are recommended in the ATLAS collaboration: HighPtCaloOnly, TightTrackOnly\_VarRad, TightTrackOnly\_FixedRad, Tight\_VarRad and Loose\_VarRad. These operating points are chosen to balance the high efficiency of identifying prompt electrons, whether they are isolated or in a busy environment, and the good rejection of fake/non-prompt electrons. Their definition is in Table~\ref{tab:ele_wps}.

\begin{table}[!ht]
\centering
\tbl{
	Definition of the electron isolation working points.
	For the calorimeter isolation, a cone size
	of $\Delta R = 0.2$ is used, and of $\Delta R_{\mathrm{max}} = 0.3$ or 0.2 for track isolation, rspectively.
	Reused with permission from Refs.~\citen{Egamma_2023}.
}
{\def\arraystretch{1.4}
\begin{tabular}{lll}
	\toprule
	Selection criteria & Calorimeter isolation  & Track isolation    \\
	\midrule
	
	HighPtCaloOnly          & $\etcs < \text{max}(0.015\times \pt, 3.5)$~GeV  & --                                      \\
	TightTrackOnly\_VarRad  & --                                                     & $\ptvcs /\pt < 0.06$                    \\
	TightTrackOnly\_FixedRad& --                                                     & $\ptvcs /\pt < 0.06$ for $\et < 50$~GeV \\
	&                                                        & $\ptcs /\pt < 0.06$  for $\et > 50$~GeV \\
	Tight\_VarRad           & $\etcs/\pt < 0.06$                                     & $\ptvcs /\pt < 0.06$                    \\
	Loose\_VarRad           & $\etcs/\pt < 0.20$                                     & $\ptvcs /\pt < 0.15$                    \\
	\bottomrule
\end{tabular}
}
\label{tab:ele_wps}
\end{table}

Two sets of isolation variables are used, calorimeter- and track- based. The raw calorimeter isolation ($E_{\mathrm{T, raw}}^{\mathrm{isol}}$) is the sum of the transverse energy of the topological clusters with positive energy that are within a cone around the electron cluster center~\cite{Egamma_201516,Egamma_201517}. It includes the electron energy ($E_{\mathrm{T,core}}$), that has to be subtracted. This is done by removing the energy of the EM calorimeter cells in a rectangular cluster of size $\Delta\eta \times \Delta\phi = 5\times7$  (in EM-middle-layer units) around the electron cluster. This method is simple and stable for both prompt and fake/non-prompt electrons, regardless of their transverse momentum and pile-up. However, it does not remove all the electron energy, so a leakage correction ($E_{\mathrm{T,leakage}}$) is applied. This correction depends on the \et\ and $|\eta|$ of the electron, and is derived from MC simulation samples of single electrons without pile-up. Moreover, a correction for the pile-up and underlying-event contribution to the isolation cone ($E_{\mathrm{T,\textrm{pile-up}}}$) is estimated from the ambient energy density, using a $Z \to ee$ data sample. 

Finally, the fully corrected calorimeter isolation is computed as:
\begin{equation}
	\etcx = E_{\mathrm{T,raw}}^{\mathrm{isol\mathrm{XX}}} - E_{\mathrm{T,core}} - E_{\mathrm{T,leakage}}(\et,\eta,\Delta R) - E_{\mathrm{T,\textrm{pile-up}}}(\eta,\Delta R),
	\label{eq:calo_iso}
\end{equation}
where $\mathrm{XX}$ refers to the size of the employed cone, $\Delta R = \mathrm{XX}/100$. Frequently, XX is set to 20.

Compared to previous measurements~\cite{Egamma_201517}, the calorimeter-based isolation has an improved pile-up correction, as detailed in Ref~\citen{Egamma_2023}. The computation of the pile-up term was based on the ambient energy density $\rho$ estimated event-by-event in $|\eta| < 1.5$ ($\rho^\text{median}_\text{central}$) and $1.5 < |\eta| < 3.0$ ($\rho^\text{median}_\text{forward}$) regions. More precisely, it was estimated as the product of $\rho$ times the isolation area, using either $\rho^\text{median}_\text{central}$ or  $\rho^\text{median}_\text{forward}$  depending on the $\eta$ position of the electron. For the new subtraction, a correction factor was introduced in this product. It accounts for the smooth dependency of the pile-up contribution on $|\eta|$, due to the inherent dependency of the pile-up and the detector effects changing as a function of $|\eta|$. This $\eta$-dependent correction factor has been extracted based on the median of the calorimetric isolation distribution without pile-up subtraction as a function of $\rho^\text{median}_\text{central}$, for two annulus isolation areas around the electron ($E\text{T}^\text{cone30}-E_\text{T}^\text{cone20}$ and $E_\text{T}^\text{cone40}-E_\text{T}^\text{cone30}$), in finely granulated $\eta$ bins. This improved methodology helped to make the calorimeter isolation efficiency more $\eta$ independent, and increase its performance at high $\eta$.

The track-based isolation variable (\ptcx) is defined as the sum of the transverse momenta of the tracks that are within a cone of size XX around the electron track. The XX is usually 20 or 30, and the tracks that belong to the electron are not counted. The ATLAS experiment has redefined the calculation of the track-based isolation variable to include secondary tracks (e.g. from photon conversions), as long as they are far from the electron track. This is done by requiring that the tracks have a $|\Delta \eta| > 0.01$ difference from the electron track. This improves the separation of prompt electrons from fake/non-prompt electrons, especially in the high $\et >100$~GeV region (by more than a factor of two, in $t \bar{t}$ MC simulation), without degrading the efficiency of the prompt electrons.

For electrons that are produced in the decay of high-momentum heavy particles, other decay products can be very close to the electron direction. Therefore, the track-based isolation is also defined with a variable cone size ($p_{\mathrm{T},\mathrm{varcone}XX}$). In this case, the cone size shrinks for larger transverse momentum of the electron. The variable cone size is given by:
\begin{equation}
	\Delta R = \min \left( \frac{10}{\et[\gev]}, \Delta R_{\mathrm{max}}\right),
\end{equation}
where $\Delta R_{\mathrm{max}}$ is the maximum cone size.

\begin{figure}[!tb]
\begin{center}
	\includegraphics[width=0.325\columnwidth]{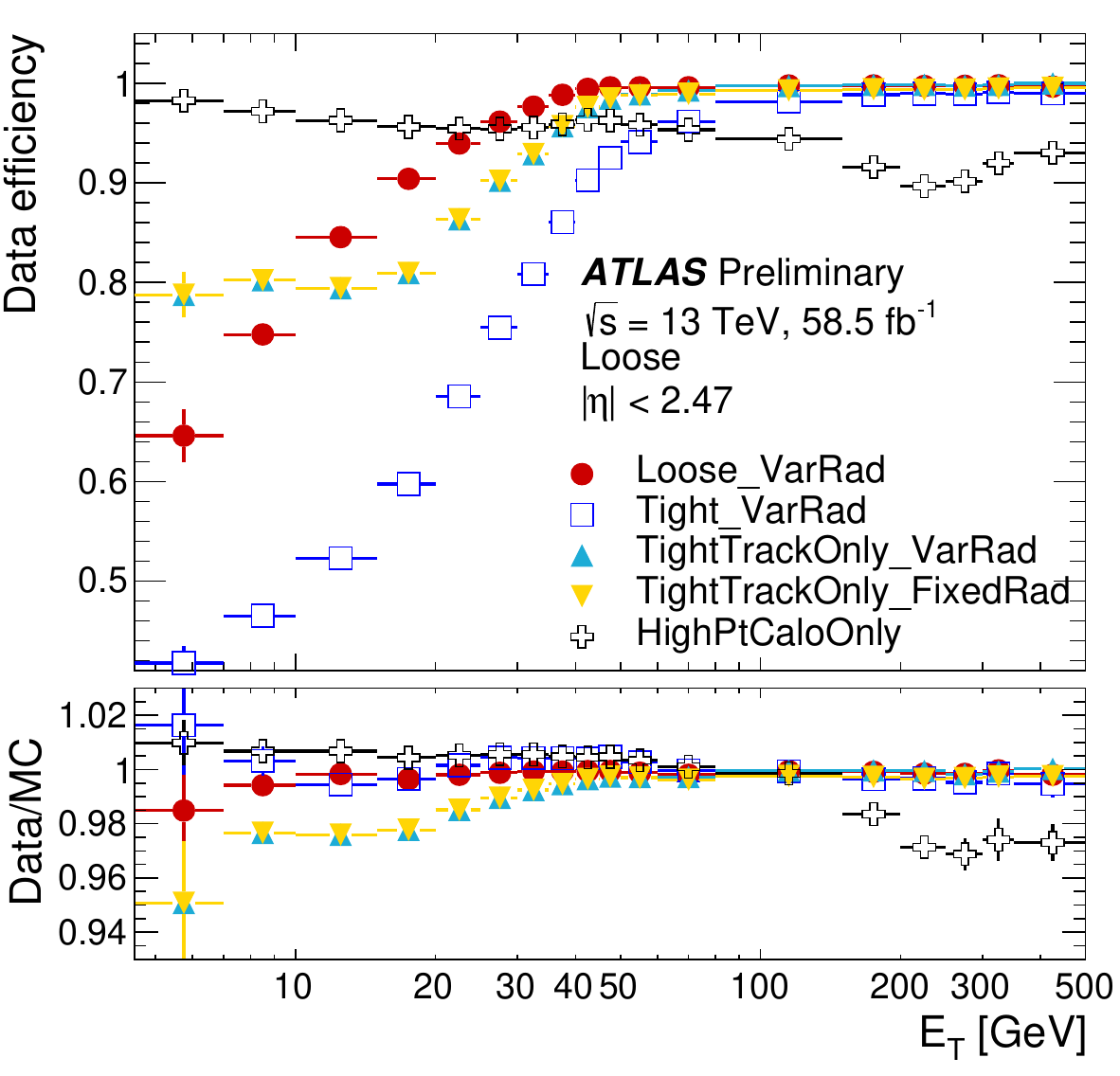}
	\includegraphics[width=0.325\columnwidth]{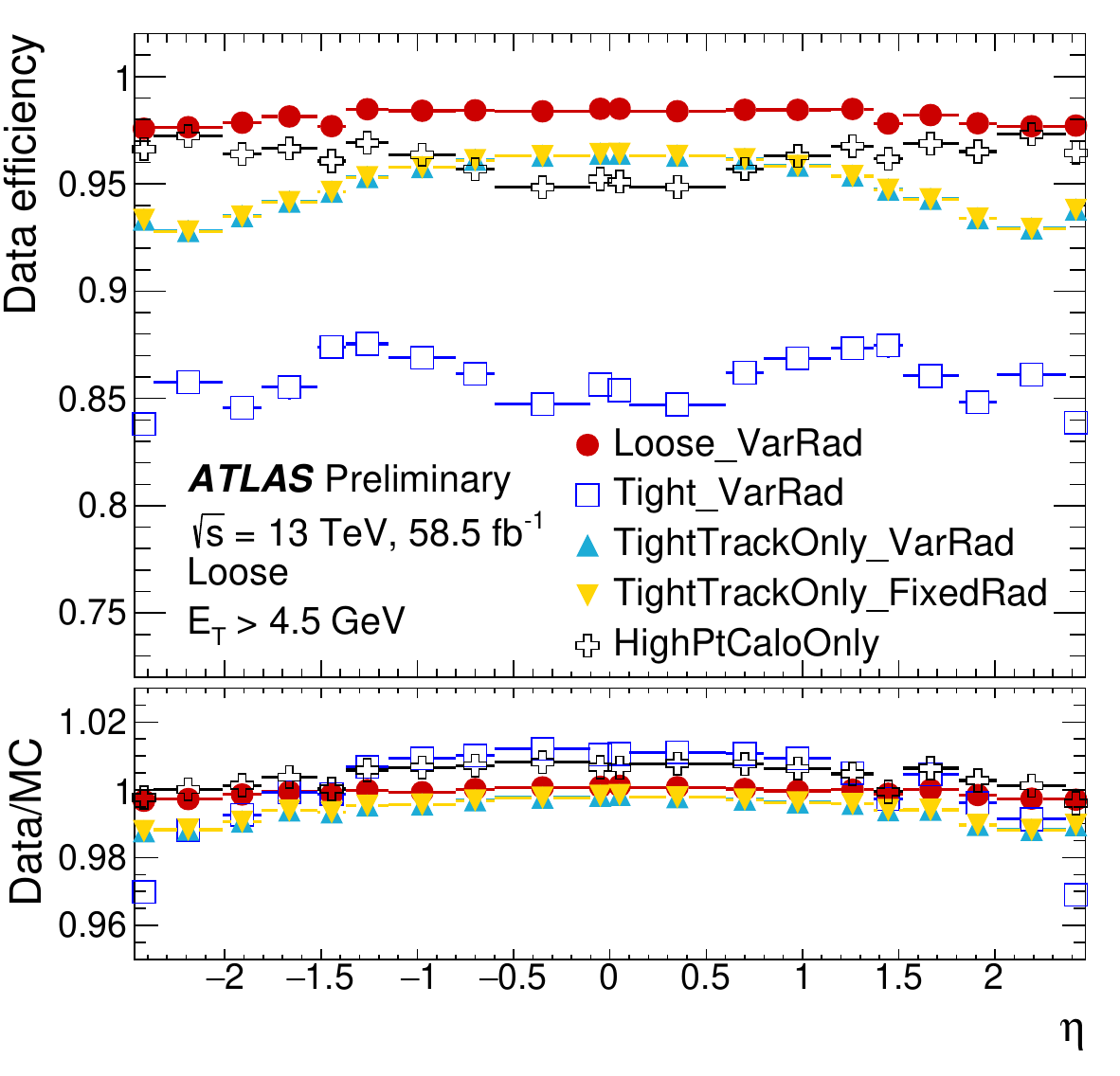}
	\includegraphics[width=0.325\columnwidth]{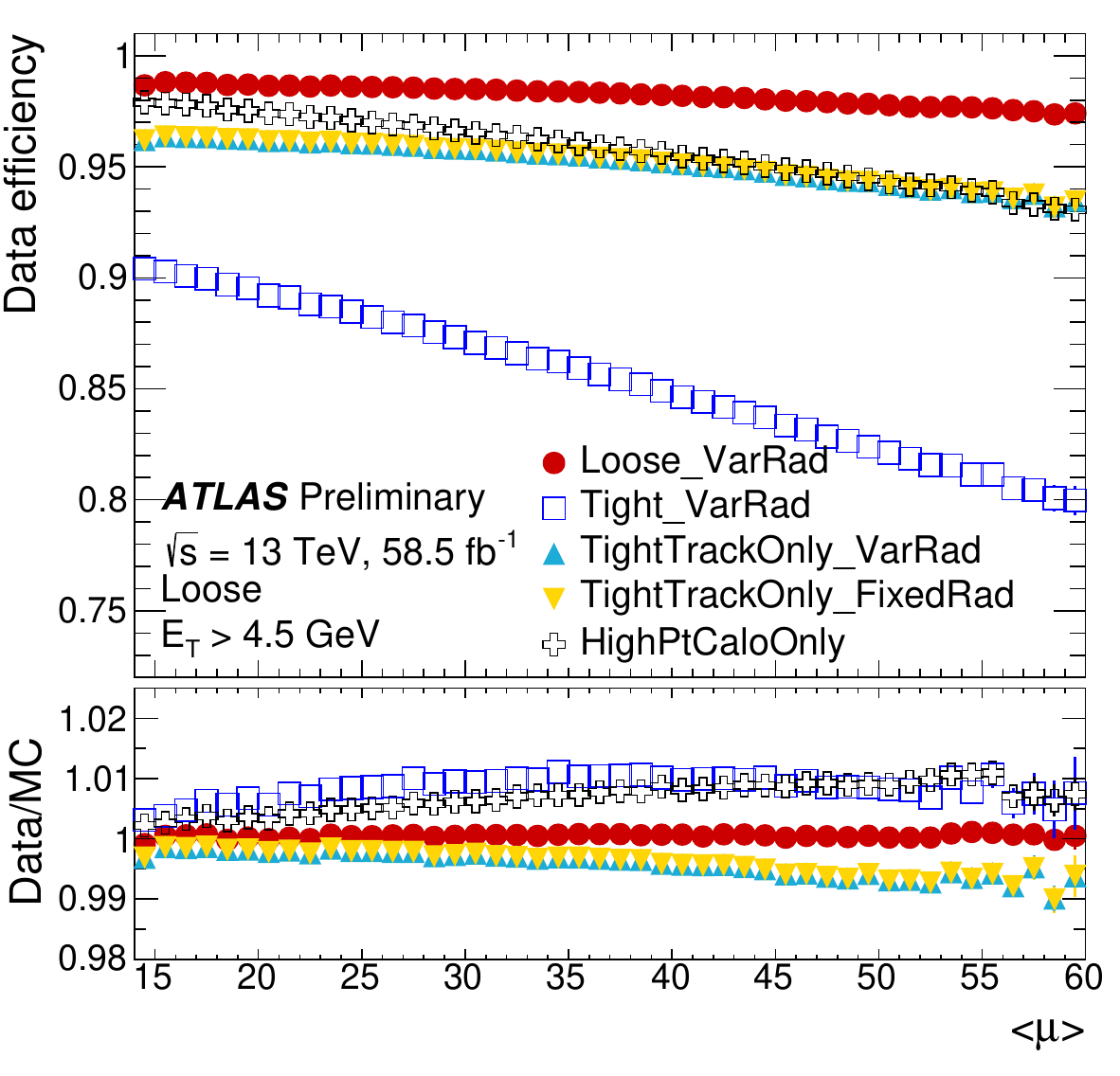}	
	\includegraphics[width=0.325\columnwidth]{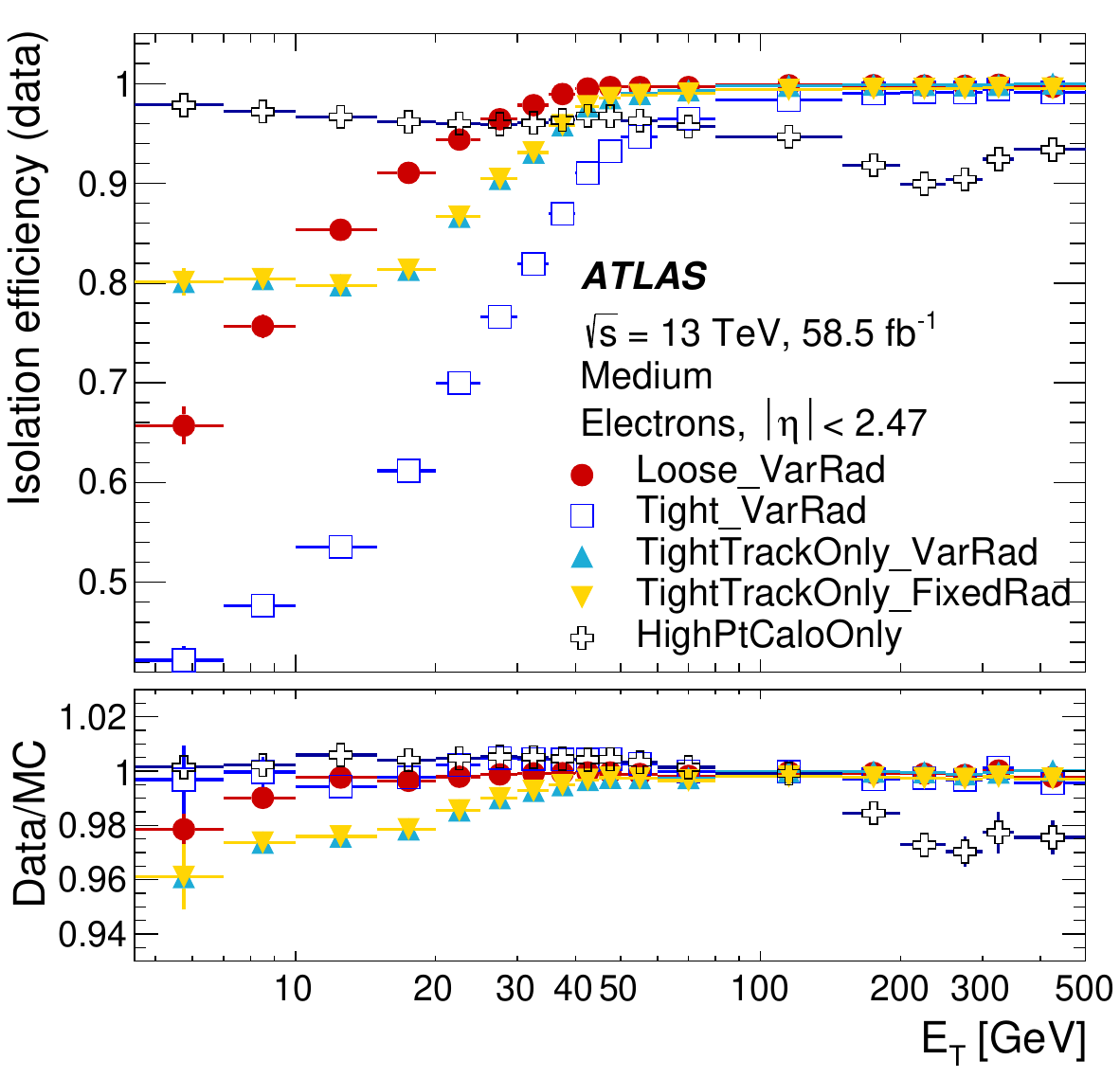}
	\includegraphics[width=0.325\columnwidth]{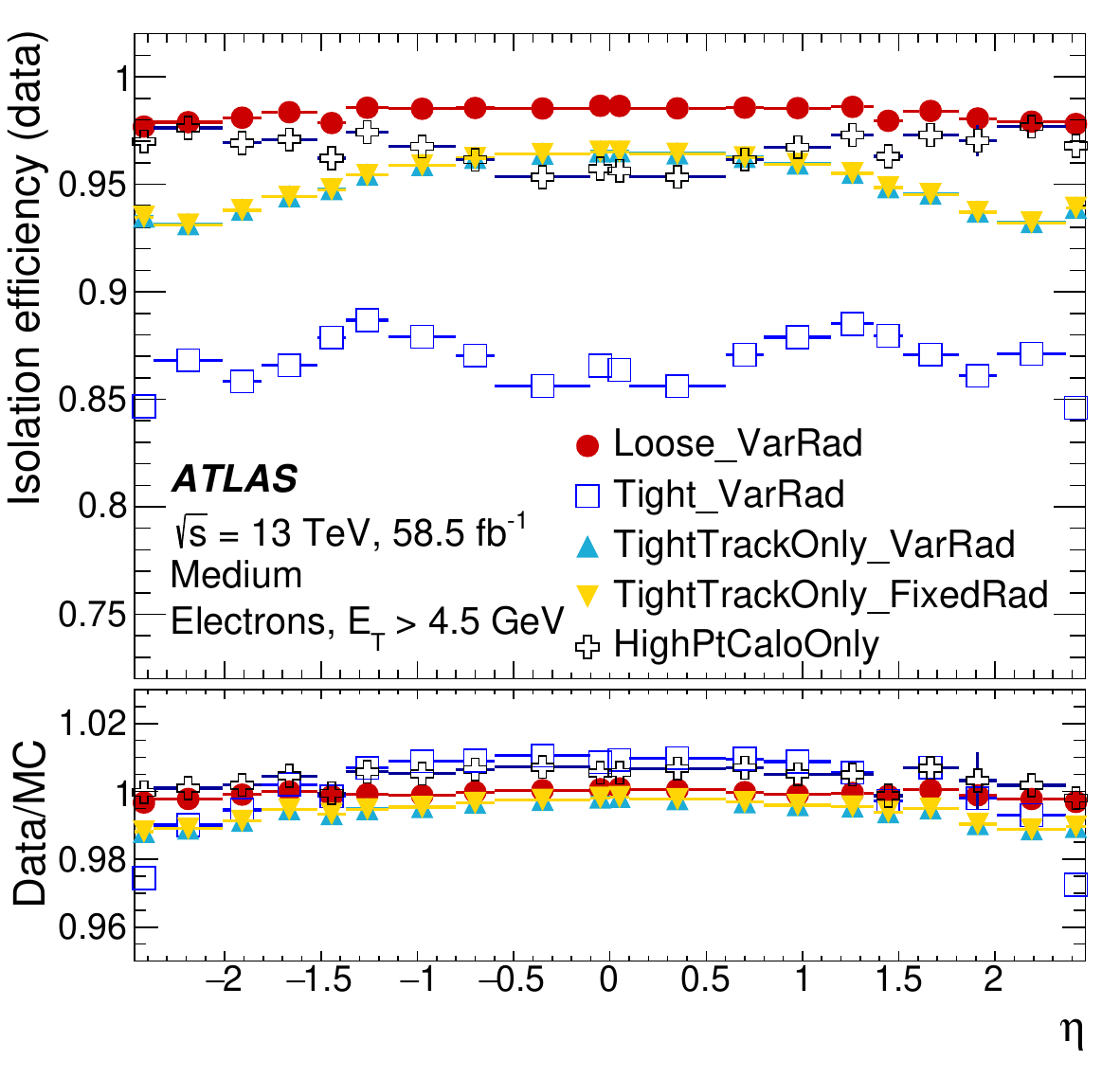}
	\includegraphics[width=0.325\columnwidth]{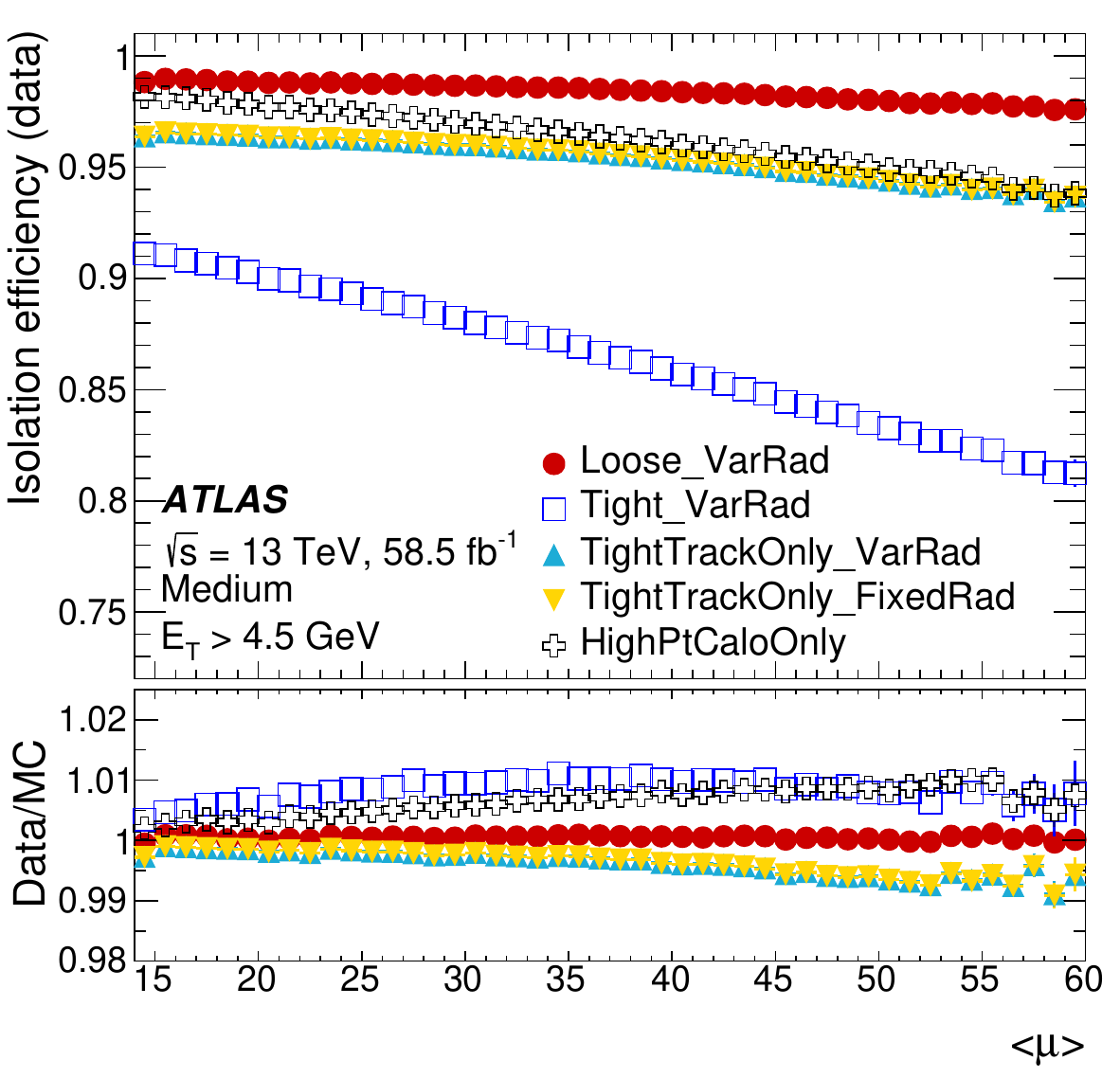}
	\includegraphics[width=0.325\columnwidth]{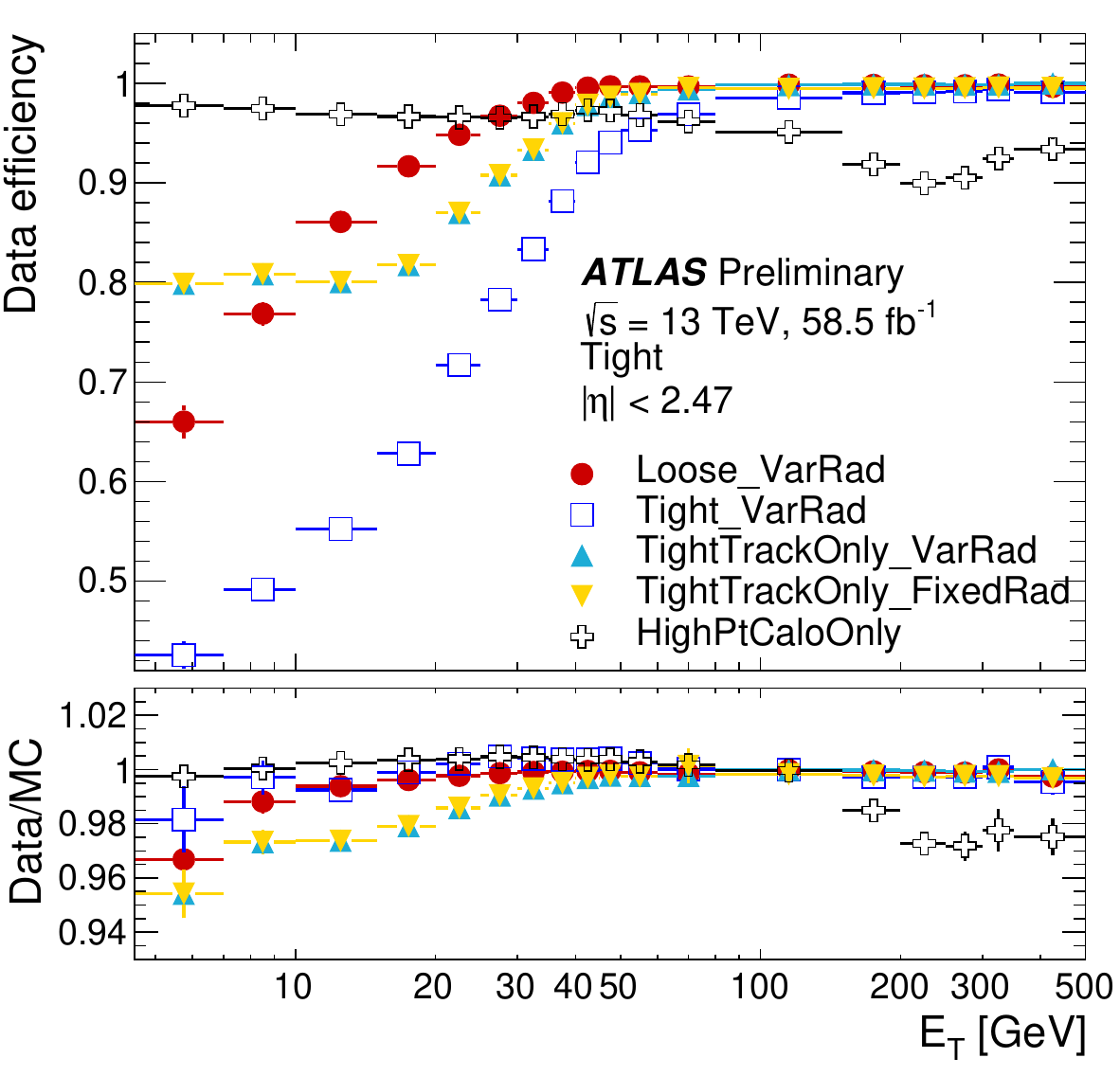}
	\includegraphics[width=0.325\columnwidth]{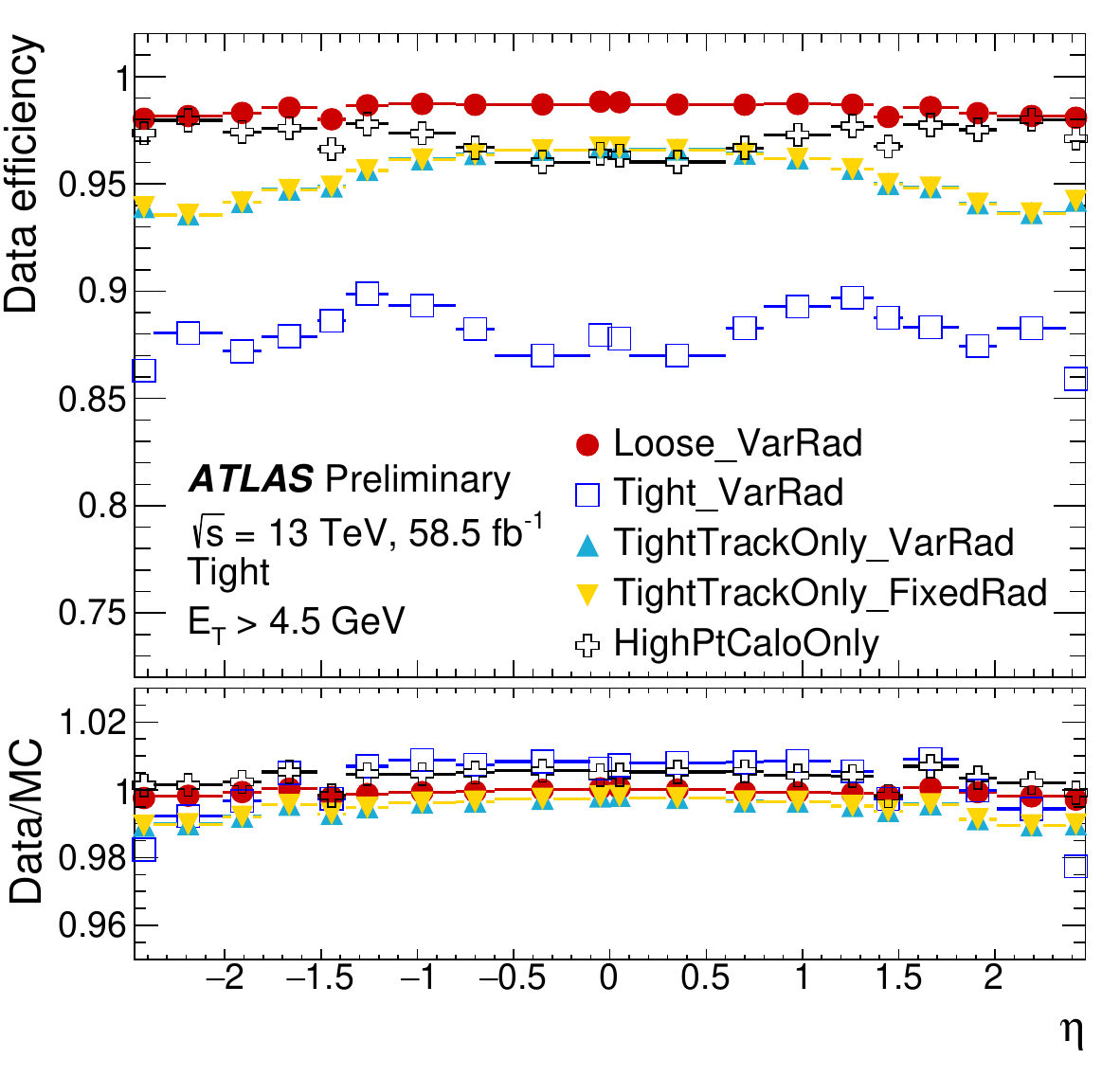}
	\includegraphics[width=0.325\columnwidth]{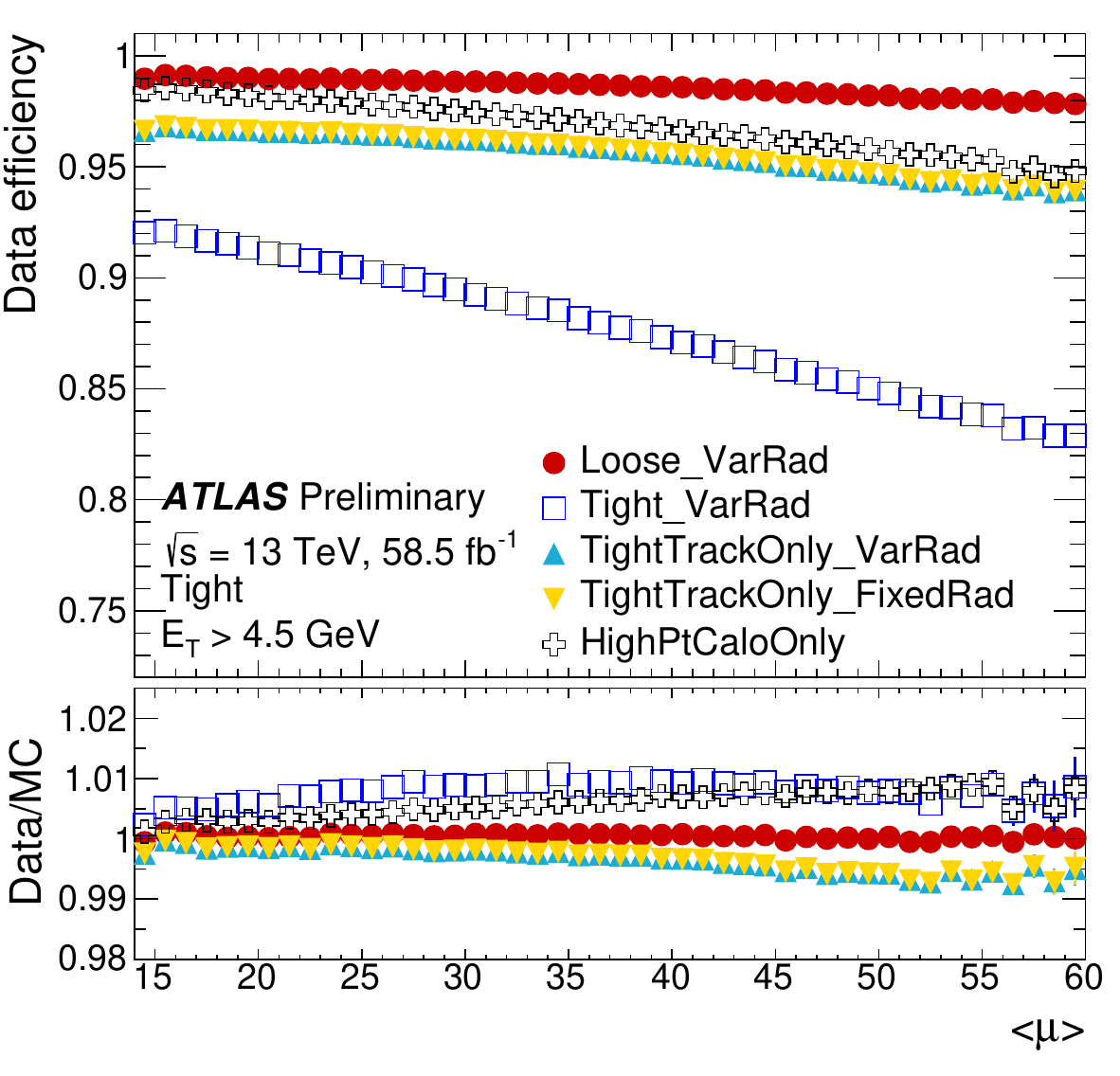}	
\end{center}
\caption{	
	Electron isolation efficiency, for Loose (top), Medium (middle) and Tight (bottom) electron identification working points. 
	Reused with permission from Refs.~\citen{Egamma_2023,Egamma_2023_PP}.
}
\label{fig:Ele_Iso_Eff}
\end{figure}

Figure~\ref{fig:Ele_Iso_Eff}, shows the data electron isolation efficiency for all five isolation working points, when the electrons are selected with the Loose, Medium or Tight identification working points. 
The electron isolation correction factors, and their associated uncertainties are also shown. The measurements are performed using the Tag\&Probe method, and $Z\to ee$ events. The isolation efficiency and the associated correction factors are found to have a dependency on the electron identification working points, even if small in $Z\to ee$ selections. This happens because the more relaxed the identification operating point is, the less isolated are the selected electrons. However, one can see significant differences between the isolation working points, especially in the low $E_T$ region, as desired. Note, here the amount of fake/non-prompt electrons is very high~\cite{Fake_Lep}.

The Tight\_VarRad isolation working point is the best at rejecting fake/non-prompt electrons below 60 GeV, and has the largest $\eta$ dependency. The HighPtCaloOnly working point is the best at rejecting fake/non-prompt electrons above 80-100 GeV in \et. The change in the efficiency at about 233 GeV in electron \et is expected, because this working point changes its definition here, as can be seen in Table~\ref{tab:ele_wps} (the $\etcs$ cut-value changes from $0.015\times p_T$ to 3.5 GeV). 
The TightTrackOnly\_VarRad and TightTrackOnly\_FixedRad working points have almost identical values for the efficiency and correction factors, as these results are obtained with a $Z\to ee $ enriched selection that has a ``clean" environment, so without too many objects around the probe electrons. If the isolation efficiency would have been obtained in a $\ttbar$ selection, or with BSM MC simulation (like $Z'$ samples, or Supersymmetric samples with boosted gluino pair production), the results would have been significantly different: the TightTrackOnly\_FixedRad efficiency would have been lower than the TightTrackOnly\_VarRad one, for electrons with $E_T > 50$~GeV. 
The Loose\_VarRad is in between the TightTrackOnly and the Tight\_VarRad isolation working points, in terms of prompt electron efficiency and fake/non-prompt background rejection, and is the most used for ATLAS analyses. It also has the most flat efficiency versus $E_T$.

The efficiencies are the same for positive and negative pseudo-rapidity values for all the isolation working points, as shown in Figure~\ref{fig:Ele_Iso_Eff}, middle.
Figure~\ref{fig:Ele_Iso_Eff}, left, shows the results are shown as a function of the average number of interactions per bunch crossing, $\muhat$. Track-based isolation variables are mostly unaffected by the additional tracks and energy deposits from pile-up collisions, because the tracks that come from pile-up vertices, or that are far from the primary vertex, are mostly discarded. The calorimeter isolation is very pile-up dependent~\cite{Egamma_201517}, and this is reflected in the isolation efficiency for the Tight\_VarRad isolation working point.  

Overall, the isolation correction factors are close to one, within 5-7\% -- a similar performance is seen also in CMS. For electrons with $E_T$ above $500$ \gev, there are not enough data events to measure the isolation efficiency, so the results from the [350, 500] \gev\ bin are used to extrapolate up to 2~TeV with uncertainties up to 40\%, depending on the isolation working point.

\section{Further improvements}
In CMS experiment, two sets of combined identification and isolation working points are optimized and recommended for the physics analyses: cut-based, and multivariate based techniques~\cite{CMS_Egamma}. 
For the isolation, they exploit the information provided by the particle-flow event reconstruction, which combines information from the inner detector and calorimeters. Thus, these variables are obtained by summing the transverse momenta of charged hadrons, photons, and neutral hadrons inside an isolation cone of size 0.3. 
Such variables are expected to perform better that the ``traditional" ones (as used by ATLAS to obtain the results presented in this document), as they are more robust against pile-up, one of the biggest challenges at the LHC.

In ATLAS, the pflow-based electron isolation working points are only experimental and studied in the context of a few data analyses, such as the ones studying the properties of the Higgs boson in the 4$\ell$  decay channel. Two examples are the Higgs boson production cross-section measurements and their EFT interpretation~\cite{Higgs_PflowIso_1}, and the search for CP violation in the decay kinematics and vector-boson production of the Higgs boson~\cite{Higgs_PflowIso_2}. Ref~\citen{Higgs_PflowIso_1}, is claiming an efficiency of 80\% for the selected isolation working point, which leads to an improvement in the efficiency by about 5\% compared with the previous analysis, for the same background rejection. 

Older studies done in ATLAS, such as this Ref.~\citen{Pflow_iso_1}, also claim significant gains when using pflow-based electron isolation working points. For example, for an efficiency of 90\%, a fake/non-prompt lepton rejection of 63\% is achieved, while the traditional isolation techniques reach only 40\%.

Given all the improvements seen so far, pflow-based electron isolation working points will probably replace some of the traditional isolation working points in the future, also in the ATLAS experiment.

\section{Conclusions}	
This document presented a general overview of the electron reconstruction, identification, and isolation performance in the ATLAS experiment, with brief comparisons with the performance seen in the CMS experiment. In ATLAS, the electron reconstruction algorithm is based on superclusters, and has an efficiency better than 95\% in the $E_T > 10$~GeV region. The electron identification is based on a likelihood discriminant, and three working points are defined and proposed to the collaboration. Their definition is a balance between a good background rejection and a high acceptance of prompt electrons. For the Tight (Loose) working point, below $50$ GeV the prompt electron identification efficiency can be as low as 70\% (88\%), increasing to $> 88\%$ (95\%) above 80 GeV. The electron isolation variables are redefined to either increase the rejection of electrons from photon conversion in the $E_T > 100$~GeV range, or to decrease the isolation efficiency dependency versus $\eta$. A total of five working points are defined and proposed, and their performance was discussed too.

\vspace{0.7cm}\hspace{-1.cm}
Copyright 2023 CERN for the benefit of the ATLAS Collaboration. CC-BY-4.0 license.
\vspace{0.7cm}




\end{document}